\documentclass[aps,prl,preprintnumbers,twocolumn,groupedaddress,nofootinbib]{revtex4}

\usepackage{latexsym}
\usepackage{amsmath}
\usepackage{amsfonts}
\usepackage{amssymb}
\usepackage{amscd}
\usepackage{amsthm}
\usepackage{placeins}
\usepackage{bm}

\usepackage{ulem}
\usepackage{graphicx}
\usepackage{color}

\usepackage[latin1,utf8]{inputenc}
\usepackage[T1]{fontenc}

\allowdisplaybreaks

\newcommand{\bq}{\begin{eqnarray}}
\newcommand{\eq}{\end{eqnarray}}

\newcommand{\Eulerconstant}{\gamma_{\mathrm{E}}}

\newcommand{\Dint}{D_{\mathrm{int}}}

\newcommand{\eps}{\varepsilon}

\newcommand{\loopnumber}{l}

\newcommand{\nedges}{N}

\newcommand{\NB}{N_B}

\newcommand{\NF}{N_F}

\newcommand{\NV}{n}

\newcommand{\ND}{N_D}

\newcommand{\laportaorder}{(a,w,o,|\mu|,\dots)}

\newcommand{\arbitraryscale}{m}

\newcommand{\Divisor}{P}
\newcommand{\divisor}{p}

\newcommand{\Baikovvariable}{\sigma}

\newcommand{\preall}{C_{\eps}}
\newcommand{\preabs}{C_{\mathrm{abs}}}
\newcommand{\prerel}{C_{\mathrm{rel}}}
\newcommand{\preclutch}{C_{\mathrm{clutch}}}
\newcommand{\prebaikov}{C_{\mathrm{Baikov}}}

\newcommand{\Fcomb}{F_{\mathrm{comb}}}
\newcommand{\Fgeom}{F_{\mathrm{geom}}}
\newcommand{\Fgen}{F}

\newcommand{\Hgeom}{H_{\mathrm{geom}}}
\newcommand{\Hgen}{H}

\newcommand{\differentialform}{\Psi}

\newcommand{\Agen}{\Omega}

\newcommand{\absmu}{|\mu|}

\theoremstyle{plain}

\newcommand{\arxivdate}{June 10, 2025}

\begin{document}

\title{\boldmath{The geometric bookkeeping guide to Feynman integral reduction and $\varepsilon$-factorised differential equations}}

\author{
The $\varepsilon$-collaboration: 
          Iris Bree${}^{a}$,
          Federico Gasparotto${}^{b}$,
          Antonela Matija\v{s}i\'c${}^{a}$,
          Pouria Mazloumi${}^{a}$,
          Dmytro Melnichenko${}^{a}$,
          Sebastian P\"ogel${}^{c}$, 
          Toni Teschke${}^{a}$, 
          Xing Wang${}^{d}$,
          Stefan Weinzierl${}^{a}$,
          Konglong Wu${}^{e}$ and
          Xiaofeng Xu${}^{a,f}$
}
\affiliation{
${}^{a}$ PRISMA Cluster of Excellence, Institut f{\"u}r Physik, Johannes Gutenberg-Universit\"at Mainz, D-55099 Mainz, Germany, \\
${}^{b}$ Bethe Center for Theoretical Physics, Universität Bonn, D-53115 Bonn, Germany \\
${}^{c}$ Paul Scherrer Institut, CH-5232 Villigen, Switzerland, \\
${}^{d}$ School of Science and Engineering, The Chinese University of Hong Kong, Shenzhen 518172, China, \\
${}^{e}$ School of Physics and Technology, Wuhan University, Wuhan 430072, China, \\
${}^{f}$ Department of Physics, Xiamen University, Xiamen, 361005, China
}

\date{\arxivdate}

\begin{abstract}
We report on three improvements in the context of Feynman integral reduction and $\varepsilon$-factorised differential equations:
Firstly, we show that with a specific choice of prefactors, we trivialise the $\varepsilon$-dependence of the integration-by-parts identities.
Secondly, we observe that with a specific choice of order relation in the Laporta algorithm, we directly obtain a basis of master integrals,
whose differential equation on the maximal cut is in Laurent polynomial form with respect to $\varepsilon$ and compatible with a particular filtration.
Thirdly, we prove that such a differential equation can always be transformed to an $\varepsilon$-factorised form.
This provides a systematic algorithm to obtain an $\varepsilon$-factorised differential equation for any Feynman integral.
Furthermore, the choices for the prefactors and the order relation significantly improve the efficiency of the reduction algorithm.
\end{abstract}

\maketitle

\section{Introduction}
\label{sect:intro}

Recent advancements in particle physics 
experiments at the Large Hadron Collider are based on precision measurements 
which in turn demand increasing precision of theoretical predictions.
Given small couplings at high--energies, perturbative quantum field theory is the suitable method 
to compute different processes. 
Thus, computing Feynman integrals is the core of any precision calculations
and techniques for these computations are a topic of current research interests \cite{DHoker:2023khh,delaCruz:2024xit,Baune:2024biq,Baune:2024ber,Jockers:2024uan,Gehrmann:2024tds,Pogel:2024sdi,Duhr:2024xsy,Gasparotto:2024bku,Duhr:2024uid,DHoker:2025szl,DHoker:2025dhv,Duhr:2025ppd,Duhr:2025tdf,Becchetti:2025oyb,Duhr:2025lbz,Chaubey:2025adn}.

An effective way to compute Feynman integrals 
is the method of differential equations \cite{Kotikov:1990kg,Kotikov:1991pm,Remiddi:1997ny,Gehrmann:1999as}.
This method can be used analytically or numerically \cite{Liu:2022chg,Liu:2017jxz,Liu:2022mfb,Hidding:2020ytt,Armadillo:2022ugh,Prisco:2025wqs,PetitRosas:2025xhm}.
One utilises integration-by-parts identities to derive a set of (non-$\eps$-factorised) differential equations \cite{Tkachov:1981wb,Chetyrkin:1981qh,Laporta:2000dsw}.
This step is algorithmic and involves only linear algebra.
The only limitation is the availability of computing resources. 
For an analytic calculation one usually performs two additional steps: 
In the second step, one transforms the system of differential equations to an $\eps$-factorised form \cite{Henn:2013pwa}.
In the last step, one solves the $\eps$-factorised differential equations order by order in $\eps$ in terms of iterated integrals \cite{Chen}.
The third step is also straightforward, and there are no conceptual issues, provided appropriate boundary values are given.
Since the boundary values depend on one kinematic variable less, they are simpler to calculate. In fact,
they can be recursively reduced to single-mass vacuum integrals \cite{Liu:2022chg,Liu:2017jxz,Liu:2022mfb}.
Analytic calculations are the method of choice for Feynman integrals depending on a small number of kinematic variables \cite{Coro:2025vgn}.
Rapidly converging series expansions in variables suggested by the $\eps$-factorised form
provide fast numerical evaluation routines for these Feynman integrals.

There are two bottlenecks within this approach:
The first bottleneck is the availability of computing resources for the required integration-by-parts reduction.
A major source of expression swell in integration-by-parts reduction 
are spurious polynomials in the denominator, which depend on the dimensional regulator $\eps$ and the kinematic variables $x$.
There are some heuristic methods which try to avoid the occurrence of this situation \cite{Smirnov:2020quc,Usovitsch:2020jrk}.

The second bottleneck is conceptual:
Can one always find a transformation to an $\eps$-factorised differential equation?
For families of Feynman integrals which evaluate to multiple polylogarithms a systematic procedure is known \cite{Moser:1959,Lee:2014ioa,Lee:2017oca}
and has been implemented in several computer programs \cite{Prausa:2017ltv,Gituliar:2017vzm,Lee:2020zfb}.
For Feynman integral families which go beyond multiple polylogarithms
we know examples where 
the transformation to an $\eps$-factorised form has been constructed \cite{Adams:2018yfj,Bogner:2019lfa,Muller:2022gec,Pogel:2022yat,Pogel:2022ken,Pogel:2022vat,Giroux:2022wav,Jiang:2023jmk,Giroux:2024yxu,Duhr:2024bzt,Forner:2024ojj,Schwanemann:2024kbg,Frellesvig:2024rea,Becchetti:2025oyb}.
In addition, there are methods that under specific (limiting) assumptions 
(for example
restriction to a specific geometry \cite{Duhr:2025lbz,Maggio:2025jel,Chen:2025hzq},
advance knowledge of the alphabet \cite{Dlapa:2022wdu}
or a good guess of the initial basis \cite{Gorges:2023zgv})
allow the construction of the required transformation.

In this letter, we report on three significant improvements:
\begin{enumerate}
\item We show that with a particular choice of pre\-factors we can trivialise the $\eps$-dependence of the integration-by-parts identities.
\item We observe that an order relation inspired by geometry in the reduction algorithm leads to a basis of master integrals, whose differential equation
is in a special form (which we call $\Fgen^\bullet$-compatible).
\item We present an algorithm to convert an $\Fgen^\bullet$-compatible differential equation into an $\eps$-factorised differential equation.
\end{enumerate}
In practical terms, these findings imply
significant efficiency improvements for integration-by-parts reduction and a systematic algorithm to obtain an $\eps$-factorised differential equation.

We are interested in dimensionally regulated Feynman integrals. We denote the dimensional regularisation parameter
by $\eps$, the kinematic variables by $x=(x_1,\dots,x_{\NB})$
and a basis of master integrals by $I=(I_1,\dots,I_{\NF})$.
The latter satisfies a system of first-order
differential equations 
\bq
\label{differential_equation}
 d I
 & = &
 \hat{A}\left(\eps,x\right) I,
 \;\;\;\;\;\;
 d=\sum\limits_{j=1}^{\NB} dx_j \frac{\partial}{\partial x_j},
\eq
where 
$\hat{A}(\eps,x)$ is a $\NF \times \NF$-matrix, whose entries are differential one-forms, rational
in $\eps$ and $x$.
The task is to find an invertible $\NF \times \NF$-matrix $R(\eps,x)$ such that the transformed basis $K = R^{-1} I$ satisfies
\bq
\label{eps_factorised}
 d K
 & = &
 \eps A\left(x\right) K,
\eq
where $A(x)$ is independent of $\eps$.
A differential equation of the form eq.~(\ref{eps_factorised}) is said to be $\eps$-factorised.
It is sufficient to focus on the maximal cut, as the required transformation off the maximal cut
can be obtained by solving a differential equation.
We therefore restrict to the maximal cut and with 
a slight abuse of notation, $\NF$ now denotes the number of master integrals on the maximal cut.

We utilise mathematical tools (twisted cohomology and Hodge theory)
which allow us to treat any Feynman integral independent of a specific geometry.
It is well known that the integrands of the master integrals can be viewed as twisted cohomology classes \cite{Mastrolia:2018uzb,Frellesvig:2019uqt}.
From Hodge theory, we borrow the concepts of filtrations \cite{Deligne:1970,Deligne:1971,Deligne:1974,Carlson,Voisin_book}.

The filtrations induce a decomposition of the master integrals on the maximal cut into smaller sets.
Furthermore, we perform integration-by-parts reduction with an ordering criterion based on the filtrations.
We observe that this leads to an intermediate basis $J$, such that the differential equation for $J$ on the maximal cut
is in a Laurent polynomial form
\bq
\label{Laurent_polynomial_form}
 d J
 & = &
 \sum\limits_{k=k_{\min}}^1
 \eps^k
 A^{(k)}\left(x\right) J,
\eq
where $A^{(k)}(x)$ is independent of $\eps$ and the occuring powers of $\eps$ are restricted by 
one of the filtrations (the precise statement is given in eq.~(\ref{refined_statement})).
The basis $J$ is related to the basis $I$ by a transformation $J=R_1^{-1} I$, where $R_1(\eps,x)$ is rational in $\eps$ and $x$.

In a second step we construct a matrix $R_2(\eps,x)$, which leads to a basis $K=R_2^{-1}J$, such that 
the differential equation for $K$ on the maximal cut is in $\eps$-factorised form.
The dependence of $R_2$ on $\eps$ is rather simple, however, it may involve transcendental functions of the kinematic variables $x$.
These transcendental functions are defined as the solution of a system of $\eps$-independent first-order differential equations.

\section{Concepts and method}
\label{sect:method}

In this section, we introduce the main concepts and outline the method.
A detailed description of the algorithm is given in a longer companion paper~\cite{Bree:2025tug}.

We consider the Feynman integrals on the maximal cut in either the democratic or a loop--by--loop Baikov representation \cite{Baikov:1996iu,Frellesvig:2017aai,Chen:2022lzr}.
Let $\nedges$ be the number of residues taken for the maximal cut and let ${\mathcal C}_{\mathrm{maxcut}}$ be the corresponding contour.
We denote the remaining Baikov variables by $(z_1,\dots,z_{\NV})$.
We introduce an arbitrary scale $\arbitraryscale$, which we use to render the Feynman integrals, the Baikov variables and the kinematic variables dimensionless.
We further denote the number of loops by $\loopnumber$, 
the number of space-time dimensions by $D$,
Euler's constant by $\Eulerconstant$,
and the inverse propagators by $\sigma_j$.
Within dimensional regularisation, we will always set $D=\Dint-2\eps$ with $\Dint \in {\mathbb Z}$.
We obtain the Baikov polynomials $p_i(z)$ from the Feynman integral with all propagators raised to the power one:
\bq
\label{Baikov_representation_unit_prop}
 \int\limits_{{\mathcal C}_{\mathrm{maxcut}}} \prod\limits_{r=1}^{\loopnumber} \frac{d^Dk_r}{i \pi^{\frac{D}{2}}} 
 \frac{1}{\prod\limits_{j=1}^{\nedges} \sigma_j}
 & \sim & 
 \int d^{\NV}z \;
 \prod\limits_{i \in I_{\mathrm{all}}} \left[ \divisor_i\left(z\right) \right]^{\alpha_i}.
\eq
It is important to notice that the exponents $\alpha_i$ are always of the form
\bq
\label{def_form_exponent_loop_by_loop}
 \alpha_i \; = \; 
 \frac{1}{2} \left( a_i + b_i \eps \right),
 & \mbox{with} &
 a_i,b_i \; \in \; {\mathbb Z}.
\eq
We define $I_{\mathrm{odd}}$ as the set of indices for which $a_i$ is odd
and $I_{\mathrm{even}}$ as the set of indices for which $a_i$ is even.

In order to capture possible singularities at infinity, we extend the affine space with coordinates $(z_1,\dots,z_{\NV})$
to projective space ${\mathbb C}{\mathbb P}^{\NV}$ with homogeneous coordinates $[z_0:z_1:\dots:z_{\NV}]$.
Let $d_i$ be the degree of $\divisor_i$ and denote by $\Divisor_i$ the $d_i$-homogenisation
\bq
 \Divisor_i\left(z_0,z_1,\dots,z_{\NV}\right)
 & = & 
 z_0^{d_i}
 \divisor_i\left(\frac{z_1}{z_0},\dots,\frac{z_{\NV}}{z_0}\right).
\eq
We further set $\Divisor_0(z_0,z_1,\dots,z_{\NV})=z_0$
and 
\bq
 a_0 & = &
 \left\{\begin{array}{ll}
 0 & \mbox{if} \;\; \sum\limits_{i \in I_{\mathrm{odd}}} d_i \;\; \mbox{even}, \\
 -1 & \mbox{if} \;\; \sum\limits_{i \in I_{\mathrm{odd}}} d_i \;\; \mbox{odd}, \\
 \end{array}
 \right.
 \nonumber \\
 b_0 & = & 
 - \sum\limits_{i \in I_{\mathrm{all}}} b_i d_i.
\eq
We can unify the notation by including the index $0$ in $I_{\mathrm{even}}$ or $I_{\mathrm{odd}}$, 
depending on $a_0$ being zero or $(-1)$, respectively.
We denote the resulting index sets by $I_{\mathrm{even}}^0$, $I_{\mathrm{odd}}^0$ and $I_{\mathrm{all}}^0$.

Within twisted cohomology, we may always move integer powers of the Baikov polynomials between the twist function 
and the rational differential $\NV$-form.
We can therefore define a ``minimal'' twist function, by requiring $a_i \in \{-1,0\}$ for all $i$:
\bq
 U\left(z_0,z_1,\dots,z_{\NV}\right)
 & = &
 \prod\limits_{i \in I_{\mathrm{odd}}^0} \Divisor_i^{-\frac{1}{2}+\frac{1}{2} b_i \eps}
 \prod\limits_{j \in I_{\mathrm{even}}^0} \Divisor_j^{\frac{1}{2} b_j \eps}.
\eq
$U(z)$ is a homogeneous function.
The even and the odd polynomials will play different roles in the following.
If an even polynomial is present in the denominator of the rational differential $\NV$-form, we may take a residue 
and reduce to a simpler problem with one Baikov variable less.
The odd polynomials define a geometry, which -- to a first approximation -- is associated with the Feynman integral
and given by
\bq
\label{def_hypersurface_square_free}
 y^2 & = & \prod\limits_{i \in I^0_{\mathrm{odd}}} \Divisor_i\left(z\right)
\eq
in a suitable weighted projective space.

The central objects are differential forms, which can be written as
\bq
\label{def_input_data}
 \differentialform_{\mu_0 \dots \mu_{\ND}}\left[Q\right]
 = 
 C \;
 U\left(z\right) \hat{\Phi}_{\mu_0 \dots \mu_{\ND}}\left[Q\right]
 \eta,
\eq
where $\mu_j \in {\mathbb N}_0$ and $\hat{\Phi}_{\mu_0 \dots \mu_{\ND}}[Q]$ is an $\eps$-independent meromorphic function in $z$, given by 
\bq
 \hat{\Phi}_{\mu_0 \dots \mu_{\ND}}\left[Q\right]
 & = &
 \frac{Q}{\prod\limits_{i \in I_{\mathrm{all}}^0} \Divisor_i^{\mu_i}}.
\eq
$Q$ is a homogeneous polynomial in the Baikov variables $z$.
$\eta$ is the standard $\NV$-form defined by
\bq
\label{def_eta}
 \eta & = & \sum\limits_{j=0}^{\NV} (-1)^{j}
  \; z_j \; dz_0 \wedge ... \wedge \widehat{dz_j} \wedge ... \wedge dz_{\NV},
\eq
where the hat indicates that the corresponding term is omitted.
The $\eps$-dependent prefactor $C$ is independent of $z$ (but may depend on $x$).
It is defined 
such that the overall normalisation of $\differentialform_{\mu_0 \dots \mu_{\ND}}[Q]$
by an $\eps$-dependent function is consistent
and
such that factors of $(\alpha_i-\mu_i)$, 
associated with differentiation of $\differentialform_{\mu_0 \dots \mu_{\ND}}[Q]$ with respect to $z$ or $x$,
are absorbed into the prefactor.
The latter property trivialises the $\eps$-dependence of the integration-by-parts identities
and is responsible for a significant efficiency improvement.
The explicit definition of the prefactor is given in the companion paper~\cite{Bree:2025tug}.
We further set
\bq
\label{def_mu}
 \left| \mu \right|
 & =
 \sum\limits_{i \in I_{\mathrm{all}}^0} \mu_i.
\eq
We denote the vector space spanned by the differential forms of eq.~(\ref{def_input_data}) by $\Agen^{\NV}_\omega$.
This is an infinite-dimensional vector space. 
It is customary to put a subscript $\omega=d\ln U$.
We obtain a finite-dimensional vector space by modding out linear relations
like integration-by-parts identities.
The resulting finite-dimensional vector space is the twisted cohomology group $\Hgen^{\NV}_{\omega}$.

We denote the vector space of Feynman integrals 
on the maximal cut modulo integration-by-parts identities by $V^{\NV}$, this is again a
finite-dimensional vector space.
There is an injective map
\bq
 \iota & : & V^{\NV} \rightarrowtail \Hgen^{\NV}_{\omega},
\eq
obtained from expressing the maximal cut of the Feynman integral in the Baikov representation.
In general, the map will not be surjective.
There are two reasons for this:
First of all, integration can lead to symmetries among Feynman integrals (elements in $V^n$), which are not symmetries of the integrands (elements in $\Hgen^n_{\omega}$).
Secondly, a polynomial $\divisor_j(z)$ with $j \in I_{\mathrm{even}}$ can simply be a factor $z_r=\Baikovvariable_l$,
where $\Baikovvariable_l$ is an uncut inverse propagator.
In this case, $\Hgen^{\NV}_{\omega}$ will also contain the integrands of the sector 
where the exponent of this inverse propagator is positive.
If this sector has additional master integrals, they will also appear in $\Hgen^{\NV}_{\omega}$.
We call such a sector a super-sector, and we include it in the analysis.
Taking these subtleties into account, we can entirely work in the space $\Hgen^{\NV}_{\omega}$ 
and convert back to $V^{\NV}$ in the end.

For the integrands defined in eq.~(\ref{def_input_data}) we have three types of linear relations:
Integration-by-parts identities, distribution identities and cancellation identities.
The integration-by-parts identities read
\bq
\label{eq_ibp}
\lefteqn{
 0 = 
 \frac{1}{\eps}
 \differentialform_{\mu_0 \dots \mu_i \dots \mu_{\ND}}\left[\partial_{z_j} Q_+\right]
} & & \\
 & &
 +
 \sum\limits_{i \in I_{\mathrm{all}}^0} 
 \differentialform_{\mu_0 \dots (\mu_i+1) \dots \mu_{\ND}}\left[Q_+ \cdot \left( \partial_{z_j} P_i \right) \right],
 \nonumber
\eq
where $Q_+$ is an $\eps$-independent homogeneous polynomial of degree $\deg Q_+ = \deg Q +1$.
The distribution identities are rather trivial and originate from writing a polynomial $Q=Q_1+Q_2$ as a sum of 
two other polynomials:
\bq
\label{eq_distribution}
 \differentialform_{\mu_0 \dots \mu_{\ND}}\left[Q\right]
 = 
 \differentialform_{\mu_0 \dots \mu_{\ND}}\left[Q_1\right]
 +
 \differentialform_{\mu_0 \dots \mu_{\ND}}\left[Q_2\right].
 \;
\eq
The cancellation identities originate from a cancellation of $P_j$ in the numerator and the denominator. They read
\bq
\label{eq_cancellation}
\lefteqn{
 \differentialform_{\mu_0 \dots (\mu_j+1) \dots \mu_{\ND}}\left[P_j \cdot Q\right]
 = } & &
 \\
 & &
 \frac{1}{\eps}
 \left(\frac{1}{2} a_j - \mu_j + \frac{b_j}{2} \eps\right)
 \differentialform_{\mu_0 \dots \mu_j \dots \mu_{\ND}}\left[Q\right].
 \nonumber
\eq
Note that the prefactor $C$ in eq.~(\ref{def_input_data}) 
has been meticulously defined to trivialise the $\eps$-dependence of the integration-by-parts identities.
In fact, we may reduce the subsystem formed by eqs.~(\ref{eq_ibp}) and (\ref{eq_distribution})
by setting $\eps=1$.
This is a significant efficiency improvement, as we have one variable less.
This is possible, because in the integration-by-parts identities eq.~(\ref{eq_ibp}) 
and the distribution identities eq.~(\ref{eq_distribution})
the explicit $\eps$-factors are synchronised with $\absmu$
and the $\eps$-dependence of the coefficients in the reduction
is therefore always monomial.

In the full system, this is, however, spoiled by the cancellation identities.
In these relations, the offending part comes from the bracket on the right-hand side of eq.~(\ref{eq_cancellation}).
Nevertheless, it is advantageous to reduce the subsystem formed by eqs.~(\ref{eq_ibp}) and (\ref{eq_distribution}) first and then
combine this reduced system with the cancellation identities of eq.~(\ref{eq_cancellation}).

To each object $\differentialform_{\mu_0 \dots \mu_{\ND}}[Q]$ as in eq.~(\ref{def_input_data}) we associate three integer numbers $(r,o,\absmu)$, where $\absmu$ has already been defined in eq.~(\ref{def_mu}).
We let $r$ to be the largest number such that the $r$-fold residue of $\differentialform^0_{\mu_0 \dots \mu_{\ND}}[Q]$ is non-zero, 
where $\differentialform^0_{\mu_0 \dots \mu_{\ND}}[Q]$ is defined by setting $\eps=0$ in the twist function.
The integer $o$ denotes the pole order of $\differentialform^0_{\mu_0 \dots \mu_{\ND}}[Q]$.
The pole order is the maximum of pole orders at individual points.
For $\alpha>0$, the pole order of $z^{-\alpha} dz$ at $z=0$ is $\lfloor \alpha \rfloor$, where $\lfloor x \rfloor$ denotes the floor function.
For normal-crossing singularities, the pole order is additive, i.e. the pole order of $dz_1/z_1 \wedge dz_2/z_2^2$ at
$(z_1,z_2)=(0,0)$ is $3$.
For non-normal-crossing singularities, we first need to perform a blow-up.

In addition, we introduce the concept of localisations:
We consider differential forms 
where $\Divisor_i$ with $i \in I_{\mathrm{even}}^0$ appears in the denominator of $\hat{\Phi}_{\mu_0 \dots \mu_{\ND}}[Q]$ (i.e. $\mu_i>0$). 
For those differential forms, we may take a residue at $\Divisor_i=0$.
We also say that we localise on $\Divisor_i=0$.
In ref.~\cite{Bree:2025tug} we discuss in detail, how the $\eps$-dependent part of the exponent of $\Divisor_i$ in the twist function
is treated when taking the residue.
The residue is then a differential $(\NV-1)$-form, and we consider the integration-by-parts identities of these forms on the variety defined by
$\Divisor_i=0$.
The explicit formulae are worked out in ref.~\cite{Bree:2025tug}
and have the property that they do not introduce algebraic extensions.
If the differential $(\NV-1)$-form has a further even polynomial in the denominator of $\hat{\Phi}_{\mu_0 \dots \mu_{\ND}}[Q]$, 
this process can be iterated. In this way we obtain $(\NV-2)$-forms, $(\NV-3)$-forms,  \dots, $0$-forms. 
We define the fourth integer number $a$ as
\bq
 a & = & \left\{
 \begin{array}{rl}
 -w, & \differentialform_{\mu_0 \dots \mu_{\ND}}[Q] \; \mbox{is a preferred candidate}\\
     & \mbox{from the localisations},\\
 0, & \mbox{otherwise.} \\
 \end{array}
 \right.
\eq
The purpose of the variable $a$ is to give preference to the master integrands from localisations 
as master integrands of the current problem.
The $a$-value can be computed recursively; the details can be found in~\cite{Bree:2025tug}.

Now we define the order relation for the Laporta algorithm on the space $\Agen^{\NV}_\omega$ as
\bq
 \laportaorder,
\eq 
where 
$w=\NV+r$, and
the dots stand for further criteria needed to distinguish inequivalent integrands.
The relation $a_1 < a_2$ implies $\differentialform_1 < \differentialform_2$, with ties broken by $w$, etc..

The three numbers $w,o,\absmu$ also define three filtrations $W_\bullet$, $\Fgeom^\bullet$ and $\Fcomb^\bullet$ on the space $\Agen^{\NV}_\omega$.
The weight filtration $W_\bullet$ is defined by 
\begin{alignat}{2}
 \differentialform_{\mu_0 \dots \mu_{\ND}}[Q] & \in W_w \Agen^{\NV}_\omega & \quad \mbox{if} & \quad \NV + r \le w.
\end{alignat}
The weight filtration is the standard weight filtration from Hodge theory.
The filtration $\Fgeom^\bullet$ is defined by
\begin{alignat}{2}
 \differentialform_{\mu_0 \dots \mu_{\ND}}[Q] & \in \Fgeom^{p} \Agen^{\NV}_\omega & \quad \mbox{if} & \quad \NV+r-o \ge p.
\end{alignat}
At fixed weight $w$, the filtration $\Fgeom^\bullet$ is a filtration by the pole order $o$.
The third filtration $\Fcomb^\bullet$ is defined by 
\begin{alignat}{2}
 \differentialform_{\mu_0 \dots \mu_{\ND}}[Q] & \in \Fcomb^{p'} \Agen^{\NV}_\omega & \quad \mbox{if} & \quad \NV-\absmu \ge p'.
\end{alignat}
The combinatorial filtration $\Fcomb^\bullet$ is a filtration by the quantity $\absmu$.
The general idea is that we always work modulo simpler terms, i.e. modulo terms with fewer residues, lower pole order
or a smaller sum of indices $\absmu$. 

We denote by $\differentialform=(\differentialform_1,\dots,\differentialform_{\NF})^T$
the basis of master integrands obtained from this algorithm.
In all examples we tested, we observed that the differential equation 
is of the form as in eq.~(\ref{Laurent_polynomial_form}).
Moreover, we always observed that if $\differentialform_i$ has $\absmu=\absmu_i$
and $\differentialform_j$ has $\absmu=\absmu_j$, then
\bq
\label{refined_statement}
 A_{ij}\left(\eps,x\right)
 & = &
 \sum\limits_{k=-(\absmu_i-\absmu_j)}^1
 \eps^k A^{(k)}_{ij}\left(x\right).
\eq
We call a differential equation which satisfies eq.~(\ref{refined_statement}) an $\Fgen^\bullet$-compatible differential equation for the filtration $\Fcomb^\bullet$.
An $\Fgen^\bullet$-compatible differential equation implies Griffiths transversality~\cite{Griffiths:1969}. 
It is, however, a stronger statement, as it requires the differential equation to be in Laurent polynomial form with restrictions on the occurring powers of $\eps$.

We now prove that we may always construct an $\eps$-factorised differential equation from an $\Fgen^\bullet$-compatible differential equation.
Let $J=(J_1,\dots,J_{\NF})^T$ be a basis with an $\Fgen^\bullet$-compatible differential equation
and assume that $J$ is ordered according to the $\Fgen^\bullet$-filtration, i.e.
$J_1 \in \Fgen^{p_{\max}} V^{\NV}$ and $J_{\NF} \in \Fgen^{p_{\min}} V^{\NV}$.
The matrix $A$ defined by
\bq
 A & = &
 \sum\limits_{k=-\NV}^1
 \eps^k
 A^{(k)}\left(x\right)
\eq
has then a block structure induced by the $\Fgen^\bullet$-filtration.
It will be convenient to organise the matrix $A$ as 
\bq
 A & = & 
 \sum\limits_{k=-\NV}^1
 B^{(k)}\left(x\right),
\eq
with $B^{(1)}(x)=\eps A^{(1)}(x)$. 
For $k<1$ the matrices $B^{(k)}(x)$ are lower block-triangular.
The blocks on the lower $j$-th block sub-diagonal are given by the terms of order $\eps^{\NV+k-j}$ of 
the corresponding blocks of $A$.
We say that a term is of $B$-order $k$ if the term appears in $B^{(k)}$.

We now construct the matrix $R_2$, leading to the $\eps$-factorised basis $K=R_2^{-1} J$.
The matrix $R_2$ is given as
\bq
 R_2 & = & R_2^{(-\NV)} R_2^{(-\NV+1)} \dots R_2^{(-1)} R_2^{(0)}.
\eq
All matrices $R_2^{(k)}$ are lower block-triangular.
The matrix $R_2^{(-\NV)}$ is of $B$-order $(-\NV)$,
the matrices $R_2^{(k)}$ with $-\NV < k \le 0$ are given by
\bq
 R_2^{(k)}
 & = &
 {\bf 1} + T_2^{(k)},
\eq
where $T_2^{(k)}$ is of $B$-order $k$, and
${\bf 1}$ denotes the $\NF \times \NF$ unit matrix.
We construct the matrices $R_2^{(k)}$ iteratively, starting from $k=-\NV$ and ending with $k=0$.
We set $\tilde{A}^{(-\NV)}=A$ and
\bq
 \tilde{A}^{(k+1)}
 =
 \left(R_2^{(k)}\right)^{-1} \tilde{A}^{(k)} R_2^{(k)} - \left(R_2^{(k)}\right)^{-1} d R_2^{(k)}.
 \;
\eq
The matrix $R_2^{(k)}$ is determined by
\bq
\label{eq_U_k}
 \left. \left[ \left(R_2^{(k)}\right)^{-1} \tilde{A}^{(k)} R_2^{(k)} - \left(R_2^{(k)}\right)^{-1} d R_2^{(k)} \right] \right|_{k}
 = 0,
 \;
\eq
where $|_k$ indicates that only terms of $B$-order $k$ are taken.
Eq.~(\ref{eq_U_k}) defines an $\eps$-independent system of first-order differential equations for the unknown functions in the ansatz for $R_2^{(k)}$.
There are as many equations as there are unknown functions.
Eq.~(\ref{eq_U_k}) also ensures that $\tilde{A}^{(k+1)}$ only has terms of $B$-order $\{k+1,\dots,1\}$.
Thus, in every iteration step, we improve the $B$-order.
After transformation with $R_2^{(0)}$, the matrix $\tilde{A}^{(1)}$ only has terms of $B$-order $1$.

This completes the construction of the $\eps$-factorised differential equation on the maximal cut.
The extension beyond the maximal cut is straightforward: Any offending term is strictly lower block triangular
and can be removed with an ansatz similar to $R_2$.

\section{Examples}
\label{sect:examples}

We have tested the method on several known examples, see also in the longer companion paper \cite{Bree:2025tug}.
In the ``End Matter''-section, we give a pedagogical example of a Feynman integral with non-trivial
filtrations.
We construct the $\eps$-factorised form without using any information on the specific geometry (an elliptic curve in this case).

\begin{figure}
\begin{center}
\includegraphics[scale=0.5]{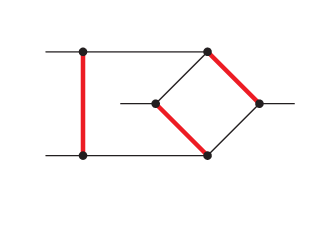}
\includegraphics[scale=0.5]{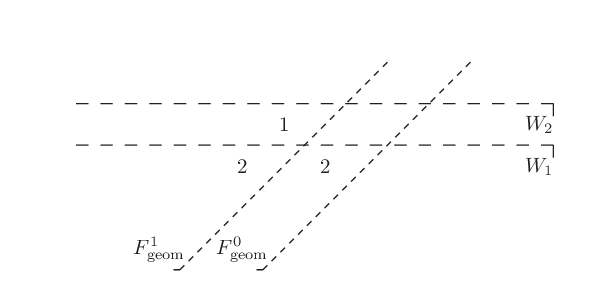}
\end{center}
\caption{
A non-planar double-box integral with internal masses (indicated by red lines).
The top sector has $5$ master integrals, which decompose
with respect to the filtrations 
as shown in the right figure.
}
\label{fig:moeller}
\end{figure}

With the method outlined in this paper, we were able to compute previously unknown Feynman integrals
through $\eps$-factorised differential equations, including the parts beyond the maximal cut.
One example is a non-planar double box integral with internal masses as indicated in fig.~\ref{fig:moeller}.
This integral contributes to M{\o}ller scattering.
It is known that the maximal cut of the top sector involves a genus two curve \cite{Marzucca:2023gto}.
The top sector has five master integrals, and the $\Fgeom^\bullet$-filtration and the $W_\bullet$-filtration decompose this sector into
$5=2+2+1$, as shown in fig.~\ref{fig:moeller}.

\begin{figure}
\begin{center}
\includegraphics[scale=0.5]{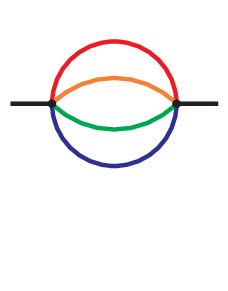}
\includegraphics[scale=0.5]{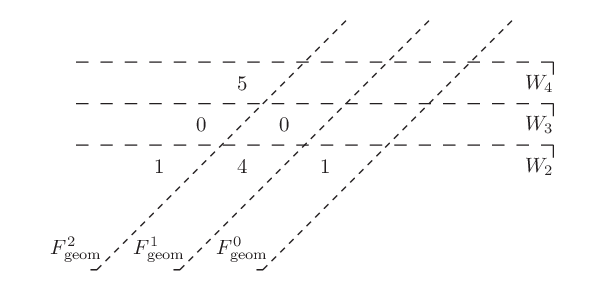}
\end{center}
\caption{
The three-loop banana graph with unequal masses.
The top sector has $11$ master integrals, which decompose
with respect to the filtrations 
as shown in the right figure.
}
\label{fig:banana}
\end{figure}

An even more advanced example is the three-loop banana graph with four unequal masses, shown in fig.~\ref{fig:banana}.
The geometry involves a $\mathrm{K}3$-surface.
The top sector consists of $11$ master integrals
and the $\Fgeom^\bullet$-filtration and the $W_\bullet$-filtration decompose this sector into
$11=1+4+1+5$, as shown in fig.~\ref{fig:banana}.

We observe significant efficiency improvements.
A measure, which is independent of the implementation of the algorithm is the size
of the differential equation. We observe for non-trivial systems
reductions in size up to a factor of $200$ after step $1$ and a further reduction in size up to a factor of $10$ after step 2.
Clearly, reconstructing with finite field methods \cite{Peraro:2016wsq,Peraro:2019svx} rational functions which are significantly smaller,
is more efficient.
The factor $200$ after step one is observed in the H-graph of ref.~\cite{Kreer:2024zzf}, 
for the non-planar double box integral in fig.~\ref{fig:moeller}
it is a factor of $25$, for the three-loop banana integral in fig.~\ref{fig:banana} it is just a factor of $1$.
The improvement is explained by the avoidance of spurious polynomials in the denominator of our algorithm.
For the three-loop banana integral the standard method does not introduce spurious polynomials, 
hence there is no significant improvement in the size of the differential equation.

\section{Conclusions}
\label{sect:conclusions}

In this letter, we reported on 
a systematic algorithm to obtain an $\eps$-factorised differential equation without relying on prior knowledge of the underlying geometry. 
It would be interesting to investigate whether the order relation proposed in this letter always leads to an $\Fcomb^\bullet$-compatible differential equation.
We do not know about a counter-example, but this is currently a conjecture.
Additionaly, we expect that the proposed method for integration-by-parts reduction yields further efficiency improvements.

\subsection*{Acknowledgements}

We thank Stefan M\"uller-Stach for useful discussions.
S.W. would like to thank the Kavli Institute for Theoretical Physics in Santa Barbara for hospitality.
This work has been supported by the Research Unit ``Modern Foundations of Scattering Amplitudes'' (FOR 5582)
funded by the German Research Foundation (DFG).
X.W. is supported by the University Development Fund of The Chinese University of Hong Kong, Shenzhen, under the Grant No. UDF01003912.
This research has received funding from the European Research Council (ERC) under the European Union’s Horizon 2022
Research and Innovation Program (ERC Advanced Grant No.~101097780, EFT4jets and ERC Consolidator Grant No.~101043686 LoCoMotive). 
Views and opinions expressed
are however those of the authors only and do not necessarily reflect those of the European Union or the European
Research Council Executive Agency. Neither the European Union nor the granting authority can be held responsible for
them.

\bibliography{biblio}
\bibliographystyle{h-physrev5}

\clearpage
\newpage

\section{End Matter}
\label{sect:supplement}

\begin{figure}
\begin{center}
\includegraphics[scale=0.49]{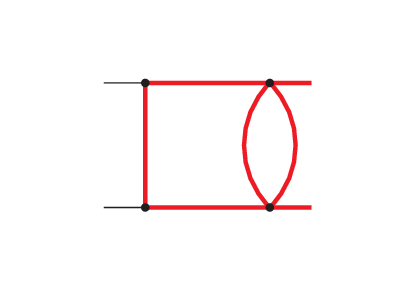}
\includegraphics[scale=0.49]{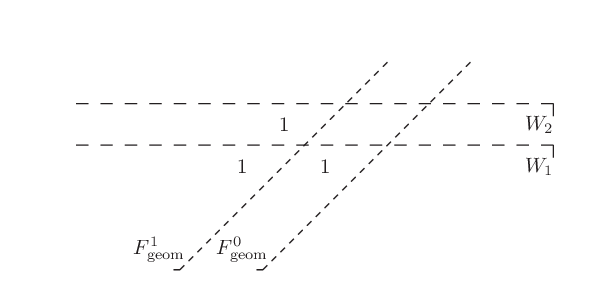}
\end{center}
\caption{
A two-loop integral with masses (indicated by red lines).
The top sector has $3$ master integrals, which decompose
with respect to the $\Fgeom^\bullet$-filtration  and the $W_\bullet$-filtration  
as shown in the right figure.
}
\label{fig:sector79}
\end{figure}

As a pedagogical example, we consider the Feynman integral named ``sector $79$'' in ref.~\cite{Muller:2022gec}.
This is one of the simplest examples with non-trivial filtrations.
There are three master integrals in the top sector.
The Feynman graph and the (final) decomposition in terms of the filtration $\Fgeom^\bullet$ and the weight filtration $W_\bullet$ are shown in fig.~\ref{fig:sector79}.
We follow the notation of ref.~\cite{Muller:2022gec}.
The inverse propagators are
\begin{align}
 \sigma_1 & = -\left(k_1+p_2\right)^2 + m^2,
 &
 \sigma_2 & = -k_1^2 + m^2,
 \\
 \sigma_3 & = -\left(k_1+p_1+p_2\right)^2 + m^2,
 &
 \sigma_4 & = -\left(k_1+k_2\right)^2 + m^2,
 \nonumber \\
 \sigma_5 & = -k_2^2,
 &
 \sigma_6 & = -\left(k_2+p_3+p_4\right)^2,
 \nonumber \\
 \sigma_7 & = -\left(k_2+p_3\right)^2 + m^2,
 &
 \sigma_9 & = -\left(k_2-p_2+p_3\right)^2,
 \nonumber \\
 \sigma_8 & = -\left(k_1+p_2-p_3\right)^2 + m^2.
 \nonumber
\end{align}
We set $x_1=s/m^2$ and $x_2=t/m^2$.
On the maximal cut we obtain, using the loop-by-loop approach, 
a minimal one-dimensional Baikov representation from an integrand with a dot on either propagator $4$ or $7$.
With $D=4-2\eps$ and $z_1=\sigma_8/m^2$ this yields
\bq
\lefteqn{
 e^{2 \eps \Eulerconstant} 
 \int\limits_{{\mathcal C}_{\mathrm{maxcut}}} \prod\limits_{r=1}^{2} \frac{d^Dk_r}{i \pi^{\frac{D}{2}}} 
 \frac{\left(\arbitraryscale^2\right)^{2+2\eps}}{\sigma_1 \sigma_2 \sigma_3 \sigma_4^2 \sigma_7}
 = 
} & &
 \\
 & &
 \prebaikov
 \int \frac{dz_1}{2\pi i} \;
 \left[ \divisor_1\left(z\right) \right]^{-\frac{1}{2}}
 \left[ \divisor_2\left(z\right) \right]^{-\frac{1}{2}-\eps}
 \left[ \divisor_3\left(z\right) \right]^{-\frac{1}{2}-\eps}.
 \;\;
 \nonumber
\eq
We have
\bq
 \prebaikov
 =  
 \frac{2^{4+4\eps} \pi^4 e^{2 \eps \Eulerconstant}}{\left[\Gamma\left(\frac{1}{2}-\eps\right)\right]^2 x_1^{1+\eps} }
 \left[\left(1-x_2\right)^2+x_1x_2\right]^\eps
\nonumber
\eq
and
\bq
 p_1 & = & z_1-x_2,
 \\
 p_2 & = & z_1+ 4-x_2,
 \nonumber \\
 p_3 & = & \left( z_1 + 1 \right)^2 - 4 \left[ x_2 +\frac{\left(1-x_2\right)^2}{x_1} \right]. 
 \nonumber
\eq
Thus $I_{\mathrm{even}}=\emptyset$ and 
$I_{\mathrm{odd}}=\{1,2,3\}$.
In this example we have $\dim V^1 = \dim \Hgen^1_\omega = 3$,
hence, we do not need to worry about symmetries introduced by integration nor about super-sectors.

The homogenisations are $P_1=z_1-x_2z_0$, $P_2=z_1+(4-x_2)z_0$ and 
\bq
 P_3 = \left( z_1 + z_0 \right)^2 - 4 \left[ x_2 +\frac{\left(1-x_2\right)^2}{x_1} \right] z_0^2.
\eq
We further introduce $P_0=z_0$ with $a_0=0$ and $b_0=6\eps$.
We have $I^0_{\mathrm{even}}=\{0\}$ and 
$I^0_{\mathrm{odd}}=\{1,2,3\}$.
The twist function is then given by
\bq
 U\left(z_0,z_1\right)
 & = &
 P_0^{3\eps}
 P_1^{-\frac{1}{2}}
 P_2^{-\frac{1}{2}-\eps}
 P_3^{-\frac{1}{2}-\eps}.
\eq
It is easy to check that with $\preabs=\eps^3 x_1$ the product $\preabs \prebaikov$ is pure of transcendental weight zero.

In $\Hgen^1_\omega$, we consider differential forms 
\bq
 \differentialform_{\mu_0 \mu_1 \mu_2 \mu_3}[Q] = \prebaikov \preall U(z) \hat{\Phi}_{\mu_0 \mu_1 \mu_2 \mu_3}[Q] \eta.
\eq
The twist function $U$ is homogeneous of degree $d_U=-2$
and $\eta = z_0 dz_1- z_1 dz_0$ is homogeneous of degree $2$, hence
$\hat{\Phi}$ has to be homogeneous of degree $0$.

We start by looking at residues.
As there is only one polynomial from the set $I^0_{\mathrm{even}}=\{0\}$, the only possibility is a residue at $P_0=0$.
For $(\mu_0,\mu_1,\mu_2,\mu_3)=(1,0,0,0)$
we must have $\deg Q=1$.
Thus, we consider
\bq
 \hat{\Phi}_{1 0 0 0}[z_1] & = & \frac{z_1}{z_0}.
\eq
In this case we have $\prerel=3\eps$ and $\preclutch=\eps^{-1}$.
At pole order $1$, there are no further possibilities to construct differential forms with non-zero residues.
Thus $\Hgeom^{(1,1)}$ is generated by
\bq
 \differentialform_2 \; = \; \differentialform_{1 0 0 0}[z_1] & = & 3 \eps^3 x_1 U(z) \frac{z_1}{z_0} \eta.
\eq
We then consider the weight $(w=1)$-part.
Within a given weight, our ordering criterion prefers differential forms of the lowest pole order.
For pole order zero, we have $\mu_0=\mu_1=\mu_2=\mu_3=0$.
In this case, we must have $\deg Q=0$ and therefore (up to irrelevant prefactors) $Q=1$.
As this is the only possibility, $\Hgeom^{(1,0)}$ is generated by
\bq
 \differentialform_1 \; = \; \differentialform_{0 0 0 0}[1] & = & \eps^3 x_1 U(z) \eta.
\eq
For the last basis element, we consider pole order one.
We have several possibilities, for example,
\bq
 \frac{z_0}{P_1},
 \;
 \frac{z_0}{P_2},
 \;
 \frac{z_0^2}{P_3}.
\eq
It will depend on the unspecified dots in the ordering criterion $\laportaorder$, which form is picked.
The actual choice is not essential here, and for concreteness, let us assume that the algorithm picks $z_0/P_1$.
With this choice we have $\prerel=-1/2$, $\preclutch=1/\eps$ and $\Hgeom^{(0,1)}$ is generated by
\bq
 \differentialform_3 \; = \; \differentialform_{0 1 0 0}[z_0] & = & -\frac{1}{2} \eps^2 x_1 U(z) \frac{z_0}{z_1-x_2z_0} \eta.
\eq
With $\differentialform=(\differentialform_1,\differentialform_2,\differentialform_3)^T$, we obtain a differential equation in Laurent polynomial form, compatible with the $\Fcomb^\bullet$-filtration:
\bq
 d \differentialform & = & \left[ B^{(1)} + B^{(0)} + B^{(-1)} \right] \differentialform,
\eq
with $B^{(-1)}$ and $B^{(0)}$ of the form
\bq
 B^{(-1)}
 & = &
 \left( \begin{array}{c|cc}
  B^{(-1)}_{11} & 0 & 0  \\
 \hline
  0 & 0  & 0  \\
  \frac{1}{\eps}B^{(-1)}_{31} & B^{(-1)}_{32} & B^{(-1)}_{33} \\
 \end{array} \right)
 \nonumber \\
 B^{(0)}
 & = & 
 \left( \begin{array}{c|cc}
  0 & 0 & 0 \\
 \hline
  B^{(0)}_{21} & 0 & 0 \\
  B^{(0)}_{31} & 0 & 0 \\
 \end{array} \right)
\eq
In these matrices, we made the $\eps$-dependence explicit and we indicated the block structure due to the $\Fcomb^\bullet$-filtration.
We then rotate the system to an $\eps$-form with the rotation matrix
\bq
 R_2 & = & R_2^{(-1)} R_2^{(0)}.
\eq
The general ansatz for $R_2^{(-1)}$ and $R_2^{(0)}$ is
\bq
 R_2^{(-1)}
 & = &
 \left( \begin{array}{c|cc}
  R^{(-1)}_{11} & 0 & 0  \\
 \hline
  \frac{1}{\eps}R^{(-1)}_{21} & R^{(-1)}_{22} & R^{(-1)}_{23} \\
  \frac{1}{\eps}R^{(-1)}_{31} & R^{(-1)}_{32} & R^{(-1)}_{33} \\
 \end{array} \right),
 \nonumber \\
 R_2^{(0)}
 & = & 
 \left( \begin{array}{c|cc}
  1 & 0 & 0 \\
 \hline
  R^{(0)}_{21} & 1 & 0 \\
  R^{(0)}_{31} & 0 & 1 \\
 \end{array} \right).
\eq
Again, we made the $\eps$-dependence explicit and we indicated the block structure due to the $\Fcomb^\bullet$-filtration.
In this particular example, we find immediately that we may set $R^{(-1)}_{21}=R^{(-1)}_{23}=0$ and $R^{(-1)}_{22}=1$.
For the other entries, we obtain a system of $\eps$-independent first-order differential equations.
An example is
\bq
 d\ln R^{(-1)}_{11}
 & = &
 B^{(-1)}_{11} + \frac{R^{(-1)}_{31}}{R^{(-1)}_{11}} B^{(1)}_{13}.
\eq
It is not too difficult to solve this system.
We then determine the preimages in $V^1$, which map to the differential forms in $\Hgen^1_\omega$.
It is sufficient to do this for $(\differentialform_1,\differentialform_2,\differentialform_3)$.
For $\differentialform_1$ and $\differentialform_2$ this is straightforward
\bq
 \iota\left( \eps^3 x_1 I_{111200100} \right) & = & \differentialform_1,
 \nonumber \\
 \iota\left( 3 \eps^3 x_1 I_{1112001\left(-1\right)0} \right) & = & \differentialform_2.
\eq
For $\differentialform_3$ one first chooses a basis $(I_1,I_2,I_3)$ in $V^1$
and determines coefficients $c_1, c_2, c_3$ from
\bq
 \iota\left(c_1 I_1 + c_2 I_2 + c_3 I_3 \right) & = & \differentialform_3.
\eq
We then find
\bq
 K_1
 & = &
 \frac{\eps^3 x_1}{R^{(-1)}_{11}} I_{111200100},
 \nonumber \\
 K_2
 & = &
 \vphantom{\frac{x^{(0)}}{x}}
 3 \eps^3 x_1 I_{1112001\left(-1\right)0} - R^{(0)}_{21} K_1.
\eq
The explicit expression for $K_3$ is slightly more lengthy, and we refrain from reporting it here.

Note that the construction presented here does not use any information on a particular geometry.
Of course, $R^{(-1)}_{11}$ is a period of an elliptic curve, but we do not need this information.
We only need to know the filtrations, but not the specific geometry.

\end{document}